\documentstyle[preprint,amssymb,prd,aps,epsfig]{revtex} %%FRK: preprint sty, also below
\tightenlines   %%FRK: for preprint, figs on 1 page

\newcommand{\bdi}{\begin{displaymath}}
\newcommand{\edi}{\end{displaymath}}
\newcommand{\bfi}{\begin{figure}}
\newcommand{\efi}{\end{figure}}

\newcommand{\beq}{\begin{equation}}
\newcommand{\eeq}{\end{equation}}
\newcommand{\beqa}{\begin{eqnarray}}
\newcommand{\eeqa}{\end{eqnarray}}

\newcommand{\rmd}{{\mathrm d}}
\newcommand{\msmall}{m}               %%CS-mass scale

\newcommand{\CS} {Chern--Si\-mons}

\newcommand {\SM}     {Standard Model}
\begin{document}
\draft

\title{Fundamental time asymmetry from nontrivial space topology}

\author{F. R. Klinkhamer\footnote{Email address:
        frans.klinkhamer@physik.uni-karlsruhe.de} }
\address{Institut f\"ur Theoretische Physik, Universit\"at Karlsruhe,
         D--76128 Karlsruhe, Germany}

\date{preprint  KA--TP--32--2001, gr-qc/0111090; Phys. Rev. D 66, 047701 (2002)}

%%FRK: comment next line out for preprint sty
%%\twocolumn[\hsize\textwidth\columnwidth\hsize\csname@twocolumnfalse\endcsname

\maketitle
\begin{abstract}
It is argued that a fundamental time asymmetry
could arise from the global structure of the space manifold.
The proposed mechanism relies on
the CPT anomaly of certain chiral gauge theories defined over a multiply
connected space manifold. The resulting time asymmetry
(microscopic arrow of time) is illustrated by a simple thought experiment.
The effect could, in principle, play a role in determining the initial
conditions of the big bang.
\end{abstract}
\vspace*{1\baselineskip}
%\keywords{time reversal; spacetime topology; electromagnetic wave propagation;
%          origin of the universe}

%%]   %%FRK: comment this line out for preprint sty

\section{Introduction}
Examining the various time asym\-me\-tries present at the macroscopic
level, Penrose \cite{P79} arrived at the following question:
``what special geometric structure did the big bang possess that
distinguishes it from the time reverse of the generic
singularities of collapse -- and why?''
He then proposed a particular condition (the vanishing of the
Weyl curvature tensor) to hold at any \emph{initial} singularity.
Whatever the precise condition may turn out to be, the crucial point is
that this condition would \emph{not} hold for
\emph{final} singularities.
This implies that the unknown physics responsible for the
initial singularity necessarily involves T, PT, CT, and CPT violation
\cite{endnoteSakurai}.

But, in that paper, Penrose did not make a concrete proposal
for the physical mechanism responsible for this
hypothetical T and CPT noninvariance. Subsequently, he has presented some
interesting, but speculative, ideas on the possible
role of quantum gravity \cite{P87}.

Here, we suggest the potential relevance of another mechanism that does not
involve gravitation directly, but does depend on the global structure
(topology) of space. The mechanism is that of the so-called CPT
anomaly \cite{K00}, which occurs for a class of chiral gauge theories
that includes
the \SM~of elementary particle physics (modulo a condition on the
ultraviolet regularization; see below).

In the remainder of this paper, we first recall the basic features of the CPT
anomaly as it applies to \SM~physics (see also Ref. \cite{K01} for a review).
We then present a thought experiment
(i.e., construct a ``clock'' $C$) that would, in principle, be able to
distinguish the initial and final singularity.
Throughout, we use natural units with $\hbar=c=1$, except when stated
otherwise \cite{endnoteFront}.

\section{Modified Maxwell theory}

It is our goal to remain as close as possible to known physics.
In addition, we prefer to give a single concrete example,
rather than to list all possibilities and confuse the reader.
We, therefore, proceed in three steps.
[In a first reading, it is possible to skip ahead to Eq. (\ref{S-MCS}), which
gives the action of the modified Maxwell theory used later on.]

First, consider the $SU(3)\times SU(2)\times U(1)$
\SM~as embedded in the $SO(10)$ gauge theory with
left-handed Weyl fermions in three spinor representations of $SO(10)$.
That is, the three families ($N_{\mathrm fam}=3$) of known left-handed quarks
and leptons, together with three hypothetical left-handed antineutrinos, are
grouped into three ${\mathbf 16}$ representations of the $SO(10)$ gauge
group \cite{Z82}. The Higgs fields are not important for our purpose.
In short, the chiral gauge theory considered has gauge group $G$ and
left-handed fermion representation $R_L$ given by
\beq \label{SO10}
(G, R_L) = (SO(10), {\mathbf 16}+{\mathbf 16}+{\mathbf 16})\;.
\eeq

Second, take the spacetime manifold ${\mathrm M}$ to be
\beq \label{R3S1}
{\mathrm M} = {\mathbb R}^3 \times S^1\;,
\eeq
with Cartesian coordinates
\beq \label{coord}
x^0\equiv c\, t,\,x^1,\,x^2\in{\mathbb R} \quad {\mathrm and}\quad x^3 \in [0,L]\;.
\eeq
The vierbeins (tetrads) are chosen to be trivial and give the Minkowski metric:
\beq \label{vierbeins}
     e_\mu^a(x) = \delta_\mu^a\,\,, \quad
     g_{\mu\nu}(x)\equiv  e_\mu^a(x)\;e_\nu^b(x)\;\eta_{ab}=\eta_{\mu\nu}\,,
\eeq
with $(\eta_{ab}) \equiv \text{diag}(-1,1,1,1)$.
Moreover, the gauge and fermion fields of the $SO(10)$ theory (\ref{SO10}) are
taken to be periodic in $x^3$ with period $L$.

Third, make the gauge-invariant ultraviolet regularization of the matter
multiplets in Eq. (\ref{SO10}) essentially the same, but with the first and
second families (i.e., the electron- and muon-type families) giving
cancelling contributions to the CPT anomaly, so that only the
contribution of the third (tau-type) family remains. For the
regularization used in Ref\cite{K00}, the odd integer
$n$ entering the anomalous term (see below) has then the value
\beq \label{nsimple}
n = \sum_{f=1}^{N_{\mathrm fam}}\, \Lambda_0^{(f)} \, /  \, |\,\Lambda_0^{(f)}|
%%= -1+1-1 = - \,1 \;,
  = +\,1-1+1 = + \,1 \;,    %%FRK: correction
\eeq
with Pauli--Villars
cutoffs  $\Lambda_0^{(f)}$ for the $x^3$-independent modes of the
fermionic fields contributing to the effective action.
For other ultraviolet regularizations, $n$ remains the sum of three odd
integers and is therefore nonzero \cite{endnoteSM}, but its value may differ
from $+\,1$.  %%FRK: correction
The ``correct'' value of the odd integer $n$
can perhaps be traced to a more fundamental theory, e.g., quantum gravity.
In this paper, we simply assume the value $n=+\,1$.  %%FRK: correction

The chiral gauge theory defined by Eqs. (\ref{SO10})--(\ref{nsimple})
turns out to have a \CS-like
term in the effective action for the $SO(10)$ gauge field, which breaks
Lorentz invariance and also T and CPT invariance. This term, which is
proportional to $n/L$, has been
discussed in great detail in Refs. \cite{K00,K01}. (Note that the
Lorentz and CPT noninvariance have also been observed in a class of
exactly solvable models in  two spacetime dimensions \cite{KN01}.)

If we now focus on the electromagnetic $U(1)$ gauge field $a_\mu(x)$
embedded in the $SO(10)$ gauge field, we have the following local terms
in the effective action at low energies \cite{K00,K01}:
\beqa \label{S-MCS}
S_{\mathrm\, MCS}^{\,{\mathbb R}^3 \times S^1}[a] &=&
 \int_{{\mathbb R}^3} \rmd x^0 \rmd x^1 \rmd x^2\, \int_{0}^{L} \rmd x^3  \;
       {\mathcal L}_{\mathrm\, MCS}\,[a]  \;,      \\[0.25cm]
{\mathcal L}_{\mathrm\, MCS}\,[a] &=&
              - \textstyle{\frac{1}{4}}\, \eta^{\kappa\mu}\,\eta^{\lambda\nu} \,
              f_{\kappa\lambda}\,f_{\mu\nu}
              - {\textstyle \frac{1}{4}}\,\msmall\,
                \epsilon^{3\lambda\mu\nu} \, f_{\lambda\mu}\,a_\nu
              \;,\label{L-MCS}
\eeqa
with the Maxwell field strength
$f_{\mu\nu} \equiv  \partial_\mu a_\nu   -   \partial_\nu a_\mu$,
the completely antisymmetric symbol $\epsilon^{\kappa\lambda\mu\nu}$
normalized by $\epsilon^{0123}= -\, 1$,     %%FRK: correction
and the \CS~mass parameter
\beq  \label{m}
\msmall          \sim \alpha/L \;,
\eeq
in terms of the fine-structure constant $\alpha \equiv e^2/(4\pi)$ and the
size $L$ of the compact dimension. The precise numerical factor in the
definition of $\msmall$ depends on the integer $n$ as given by Eq.
(\ref{nsimple}) and also on the details of the unification
and the running of the coupling constant.

The effective action (\ref{S-MCS}) describes the propagation of
electromagnetic waves \emph{in vacuo},
taking into account the effects of virtual fermions
[i.e., those of the chiral $SO(10)$ theory]. But the reflection of light by a
mirror is still described by the usual interactions of
quantum electrodynamics, at least to leading order in $\alpha$.

\section{Circularly polarized light pulses}

The propagation of light according to the  Maxwell--\CS~(MCS) theory
(\ref{L-MCS}) has been studied classically in Ref. \cite{CFJ90}
and quantum mechanically in Ref. \cite{AK01}.
Here, we are primarily interested in the classical propagation of
pulses of circularly polarized light.

Specifically, we consider light pulses propagating approximately along the
$x^2$ axis, that is, with wave vector $\vec k \equiv (k_1, k_2, k_3)$ obeying
\beq  \label{k}
k_1=0 < \msmall \ll 2\pi/L \ll  |k_3| \ll |k_2| \equiv \overline{k} \;,
\eeq
with the fine-structure constant
$\alpha$ ($\sim m L$) considered to be parametrically small.
The corresponding group velocities for left- and right-handed wave
packets have been calculated in Ref. \cite{K01}, where also the dispersion
relation can be found.

For wave vectors (\ref{k}), the magnitudes of the different group velocities
$\vec{v}_{\mathrm g}(\vec{k})$ of electromagnetic waves \emph{in vacuo} are
\begin{mathletters}
\beqa
\left|\vec v^{\,R}_{\mathrm g}(0, |k_2|,  |k_3|)\right|  &=&
\left|\vec v^{\,L}_{\mathrm g}(0,-|k_2|,- |k_3|)\right|  \nonumber\\
&\approx& 1 - \left( m^2/\,k_2^2 \right)\! \left(  1- m/\,|k_3| \right)/\,8 \,,
\!\label{vgx2R} \\[0.25cm]
\left|\vec v^{\,L}_{\mathrm g}(0, |k_2|,  |k_3|)\right|  &=&
\left|\vec v^{\,R}_{\mathrm g}(0,-|k_2|,- |k_3|)\right|  \nonumber\\
&\approx& 1 - \left( m^2/\,k_2^2 \right)\! \left(  1+ m/\,|k_3| \right)/\,8\,,
\!\label{vgx2L}
\eeqa
\end{mathletters}
up to terms of order $m^4$ and with the suffixes $L$ and $R$ indicating
left- and right-handed circular polarization. For the MCS theory (\ref{L-MCS}),
the group velocity is, in general, less or equal to $1$. Moreover, the front
velocity $v_{\rm f}\equiv \lim_{|\vec k| \to\infty} |\vec v_{\rm phase}|\,$
is $1$ in all directions and defines $c\,$; see Ref.  \cite{AK01}.

For future reference, we mention that circularly polarized light
pulses traveling along the $x^3$ axis (which corresponds to the compact dimension
of our spacetime manifold ${\mathrm M}$) have equal group velocities:
\beqa \label{vgx3}
\left|\vec v^{\,L,R}_{\mathrm g}(0, 0, k_3)\right|  &=&
|k_3|\,/\,\sqrt{k_3^2 + m^2 /4}   \;.
\eeqa
We are now ready to construct our  ``clocks.''

\section{Two clocks}

The type of clock we have in mind is a simple variation of the
``light-clock'' discussed in Ref. \cite{F63}, for example.
Our first clock $C$ consists of a single pulse of circularly
polarized light reflecting between two heavy mirrors,
firmly bolted to a common support and placed inside a vacuum chamber.
The two mirrors, $M_1$ and $M_2$, are parallel to each other and
separated by a fixed distance $D$ in the $x^2$ direction [actually, with
a small displacement in the $x^3$ direction, so as to give the wave vectors
(\ref{k}) from above]; see Fig. 1a.

The source (not shown in Fig. 1a)
produces a right-handed light pulse moving towards the right, that is,
in the positive $x^2$ direction. The pulse then oscillates between the
mirrors $M_1$ and $M_2$.
(See, e.g., Ref. \cite{BW97} for a discussion of the reflection of polarized light.)
The ``ticks'' of the clock now correspond to the
light pulse bouncing off the mirror $M_1$. With each reflection the pulse
loses some energy, which is picked up and amplified by an unspecified device.
The spacetime diagram corresponding to clock $C$ is shown in Fig. 2a. For the
MCS theory (\ref{L-MCS}), the ticks of the clock $C$ are given by ($c\equiv 1$)
\beq \label{Dt}
\Delta t \approx 2\,D\,
\left[  \,1 + \left( m^2/\,\overline{k}^{\, 2} \right)
\! \left(  1- m/\,|k_3| \right)/\,8 \,\right]\;,
\eeq
according to Eq. (\ref{vgx2R}) for $|k_3|$ and
$|k_2| \equiv \overline{k}$ as defined by Eq. (\ref{k}).

We also construct a time-reversed copy $C^\prime$ of the original clock $C$,
that is, with all motions reversed.
(See, e.g., Ref. \cite{S87} for a discussion of the time reversal transformation.)
Concretely, the source of clock $C$ is turned around
(and, if necessary, the aperture modified), so that the initial
right-handed pulse starts off to the left.
The precise nature of the mirrors in the clock $C^\prime$ is relatively
unimportant for the effect we are after and we simply consider them to be the
same as those of the clock $C$ \cite{endnoteCprime}.
Clock $C^\prime$ is shown in Fig. 1b and the
corresponding spacetime diagram in Fig. 2b.
According to Eq. (\ref{vgx2L}) for $|k_2| \equiv \overline{k}$,
the light pulse in clock $C^\prime$ travels slower than the one in $C$,
so that the ticks are longer,
\beq \label{Dtprime}
\Delta t^\prime \approx 2\,D\,
\left[  \,1 + \left( m^2/\,\overline{k}^{\,2} \right)\! \left(  1+ m/\,|k_3| \right)/\,8 \,\right]
> \Delta t\;,
\eeq
provided the \CS~mass parameter $m$ is nonzero and positive;
cf. Eqs. (\ref{S-MCS})--(\ref{m}).

Note that if both clocks $C$ and $C^\prime$ are turned by $90^\circ$ around
the $x^1$ axis (so that the initial light pulses travel exactly along the
$x^3$ axis, but in opposite directions),
the ticks become equal, according to Eq. (\ref{vgx3}).
The resulting clock $\overline{C}$ has ticks given by
\beq \label{Dtbar}
\overline{\Delta t} \approx 2\,D\,
\left[  \,1 + \left( m^2/\,\overline{k}^{\,2} \right)/\,8 \,\right]\;,
\eeq
for $|k_3| \equiv \overline{k} \gg m$ and up to terms of order $m^4$.
The behavior of clock $\overline{C}$ does not depend on the direction
($k_3 = \pm\, \overline{k}$)
of the initial right-handed pulse and is therefore invariant under time reversal
(i.e., motion reversal). But the fact remains that the original two
clocks $C$ and $C^\prime$, in the position shown in Fig. 1, would run
differently for the MCS theory (\ref{L-MCS}) \cite{endnoteCPLEAR}.

\section{Big bang vs. big crunch}

The clocks $C$ and $C^\prime$ provide an alternative to
the ones discussed implicitly by Aharony and Ne'eman \cite{AN70},
which were based on the behavior of the
$K^0-\bar{K}^0$ system with hypothetical CPT violation. As shown by these
authors, the $K^0-\bar{K}^0$ system (with nonzero CPT-violating parameter
$\delta$) could distinguish between an expanding universe and the
time-reversed copy (i.e., a contracting universe), even if the
definition of matter-antimatter was left open. The same holds for
our clocks  $C$ and $C^\prime$ (Figs. 1a and 1b),
as long as the matter content of the universe is described by an appropriate
chiral gauge field theory like the one of Eq. (\ref{SO10}) and
the space manifold is multiply connected \cite{endnoteKS}.

Following Ref. \cite{AN70}, consider a hypothetical time-symmetric universe,
now with the spacetime topology ${\mathbb R} \times S^2  \times S^1 $
(cf. Ref. \cite{endnoteKS}). Take the time interval $I=[0,\Delta\tau]$
just after the  big bang ($t=0$)
and an equal time interval $I^\prime=[\tau-\Delta\tau,\tau]$
just before the big crunch ($t=\tau$), as determined by the use of our
reference clock $\overline{C}$ or some other standard clock.
Clock $C$ running over the time interval $I$
and the time-reversed copy of clock $C$
(i.e., clock $C^\prime$) running over the time interval $I^\prime$
would then give a different number of ticks \cite{endnoteFixedn}.
Therefore, the physics near the initial singularity and the physics near the
final singularity would be different, even if the final singularity were
a time-reversed and time-translated copy of the initial
singularity. This fundamental time asymmetry (microscopic arrow of time)
is precisely one of the ingredients of the
new physics discussed by Penrose \cite{P79}.

Of course, we do not claim that the CPT anomaly necessarily plays a
role in distinguishing the big bang singularity of our own universe. After all,
we do not know for sure that the actual spacetime manifold is multiply
connected (the topology of the spacetime manifold could very well be
${\mathbb R}^4$ or ${\mathbb R}\times S^3$).  But,
in principle, the large-scale structure of spacetime could play a role
in determining the fundamental time asymmetry of the initial singularity.

\section*{Acknowledgments}
It is a pleasure to thank C. Adam for comments on the manuscript and
J. Schimmel for help with the figures.

\begin{figure}\vspace{0.5cm}
\centerline{\psfig{figure=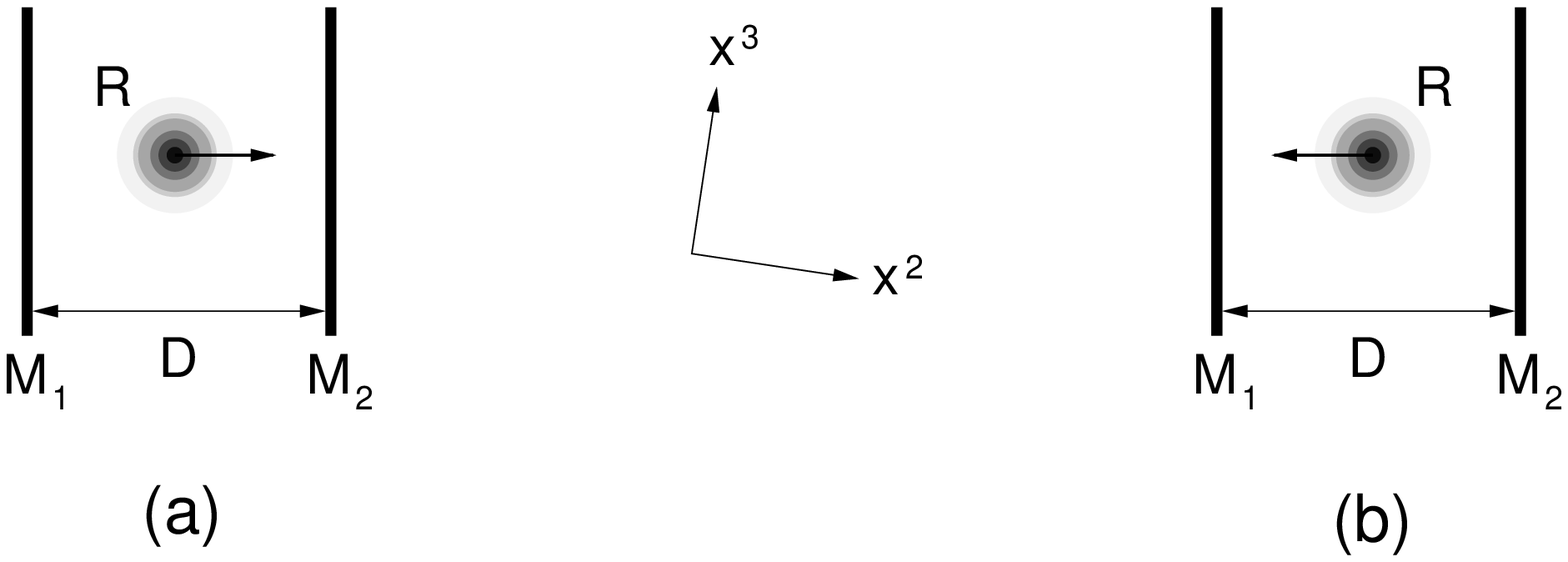,width=0.75\hsize}}\vspace{0.5\baselineskip}
\caption{(a) Sketch of clock $C$, which consists of a single pulse of circularly
polarized light reflecting between two parallel mirrors, $M_1$ and $M_2$, separated
by a fixed distance $D$ approximately  in the $x^2$ direction.
Shown is the time at which the clock is started,
with a right-handed ($R$) light pulse moving towards the right. The nonzero
energy density of the pulse is indicated by the shaded area (the contours
need not be circular).
(b) Sketch of clock $C^\prime$, which has all motions reversed compared to
clock $C$ (i.e., clock $C^\prime$ is the  time-reversed
copy of  $C$; see the main text).
Clock $C^\prime$ starts with a right-handed light pulse moving towards the left.}
\label{fig1}
\end{figure}
%%\newpage
\vspace{0.5cm}
\begin{figure}
\centerline{\psfig{figure=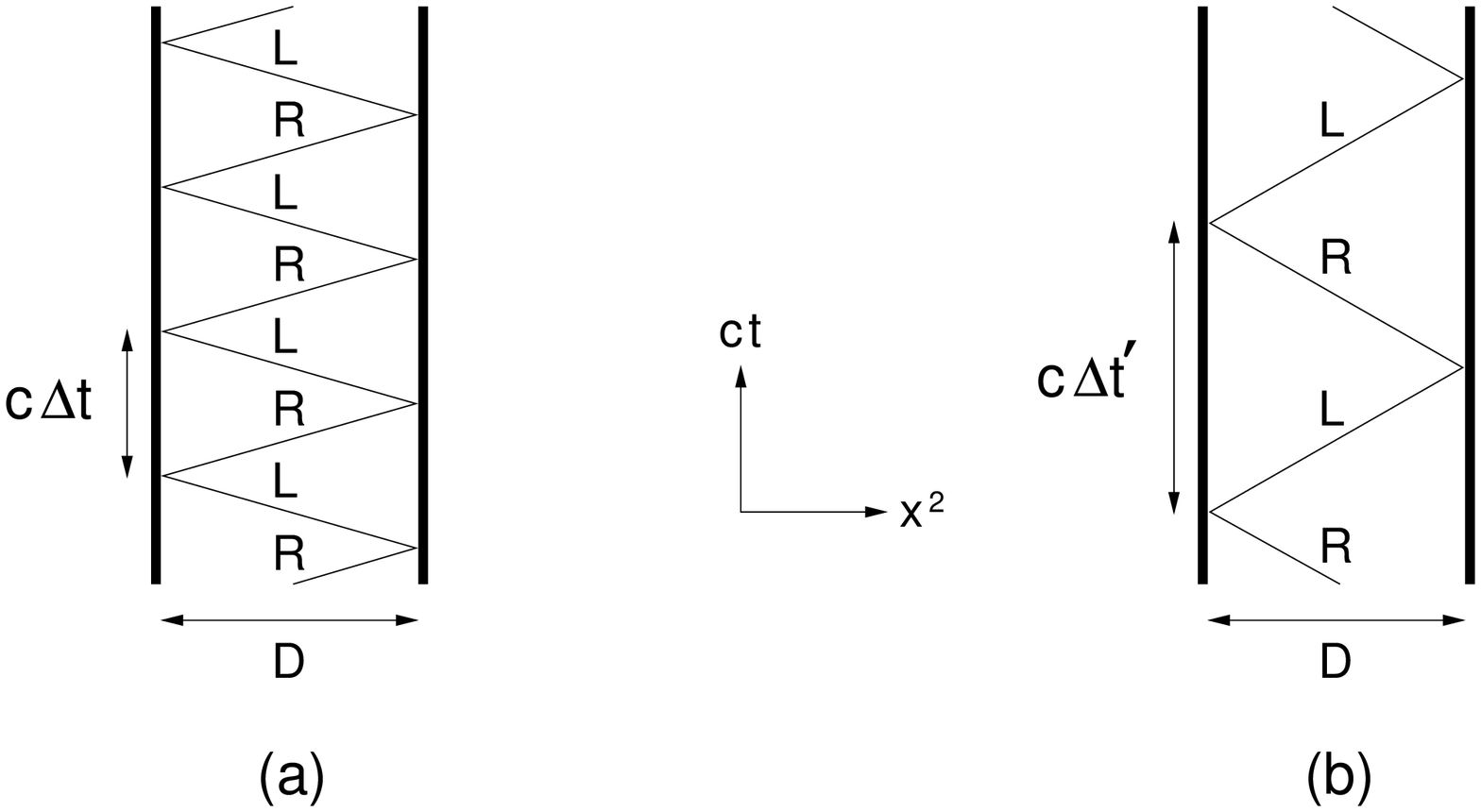,width=0.75\hsize}}\vspace{0.5\baselineskip}
\caption{(a) Schematic spacetime diagram of clock $C$ in the
Maxwell--\CS~theory (\ref{L-MCS}), with ticks $\Delta t$
between the successive reflections of the light pulse
and labels $L$ and $R$ indicating left- and right-handed circular polarization.
(The slight rotation of the mirrors of clock $C$ in Fig. 1 has been neglected here.)
The velocity $c$  is the front velocity of light; see below Eq.
(\ref{vgx2L}) in the main text. (b) Spacetime
diagram of clock $C^\prime$, with ticks $\Delta t^\prime$.}
\label{fig2}
\end{figure}


\begin{references}
\bibitem{P79}
R. Penrose, in \emph{General Relativity: An Einstein Centenary Survey}, edited
by S. W. Haw\-king and W. Israel (Cambridge University Press, Cambridge,
England, 1979), Chap. 12.
\bibitem{endnoteSakurai}
See Sec. 12.4 of Ref. \cite{P79}. Recall that C, P, and T stand for the
charge conjugation, parity reflection and time reversal transformations,
respectively; see, for example,
J. J. Sakurai, \emph{Invariance Principles and Elementary
Particles} (Princeton University Press, Princeton, NJ, 1964).
\bibitem{P87}
R. Penrose, in \emph{Three Hundred Years of Gravitation}, edited by
S. W. Haw\-king and W. Israel (Cambridge University Press, Cambridge, England, 1987),
Chap. 3.
\bibitem{K00} F.~R.~Klinkhamer, %``A CPT anomaly,''
Nucl.\ Phys.\ B {\bf 578}, 277 (2000). %%CITATION = HEP-TH 9912169;%%
\bibitem{K01}
F.~R.~Klinkhamer, in
\emph{Proceedings Seventh In\-ter\-na\-tion\-al Wigner Symposium},
edited by Y.~S.~Kim and M.~E.~Noz (in press),
%%to appear on  %%available online \texttt{http://www.physics.umd.edu/robot/},
\texttt{hep-th/0110135}. %%CITATION = HEP-TH 0110135;%%
\bibitem{endnoteFront}
Physically, $c$ corresponds to the front velocity of light,
at least for the particular modification of Maxwell theory considered;
see below Eq. (\ref{vgx2L}) in the main text.
\bibitem{Z82}
See also the discussion on $SO(10)$ grand-unified theories
in \emph{Unity of Forces in the Universe}, edited by A. Zee
(World Scientific, Singapore, 1982), Chap. 4.
\bibitem{endnoteSM}
The $SO(10)$ model considered necessarily displays the CPT anomaly,
whereas the \SM~with $N_{\mathrm fam}=3$ may or may not have the anomaly,
depending on the
type of ultraviolet regularization. The crucial point is the larger gauge
group, $SO(10) \supset SU(3)\times SU(2)\times U(1)$, which restricts
the allowed regularizations of the fermions (the gauge invariance is to
remain exact). See Refs. \cite{K00,K01} for further details.
\bibitem{KN01} F.~R.~Klinkhamer and J.~Nishimura,
%``CPT anomaly in two-dimensional chiral U(1) gauge theories,''
%%CITATION = HEP-TH 0006154;%%
Phys.\ Rev.\ D {\bf 63}, 097701 (2001);
F.~R.~Klinkhamer and  C.~Mayer,
%``Torsion and CPT anomaly in two-dimensional chiral U(1) gauge theory,''
%%CITATION = HEP-TH 0105310;%%
Nucl. Phys. {\bf B 616}, 215 (2001).
\bibitem{CFJ90} S.~M.~Carroll, G.~B.~Field, and R.~Jackiw,
%``Limits On A Lorentz And Parity Violating Modification Of Electrodynamics,''
Phys.\ Rev.\ D {\bf 41}, 1231 (1990).
%%CITATION = PHRVA,D41,1231;%%
\bibitem{AK01} C.~Adam and F.~R.~Klinkhamer,
%``Causality and CPT violation from an Abelian Chern-Simons like term,''
Nucl.\ Phys.\ B {\bf 607}, 247 (2001). %%CITATION = HEP-PH 0101087;%%
\bibitem{F63}
R.~P.~Feynman, R.~B.~Leighton, and M.~Sands,
\emph{The Feynman Lectures on Physics}
(Addison-Wesley, Reading, MA, 1963), Vol. I, Sec. 15.4.
\bibitem{BW97}
M. Born and E. Wolf, \emph{Principles of Optics\/}, 7th ed.
(Cambridge University Press, Cambridge, England, 1999), Secs. 1.4, 1.5, 14.2.
For an elementary discussion in terms of photons, see also
Vol. III, Secs. 17.4, 18.2 of Ref. \cite{F63}.
\bibitem{S87}
R.~G.~Sachs,
\emph{The Physics of Time Reversal} (Chicago University Press, Chicago, 1987);
see also Sakurai in Ref. \cite{endnoteSakurai}.
\bibitem{endnoteCprime}
For the different spacetime manifold
${\mathrm M} = {\mathbb R} \times S^1 \times S^1 \times S^1$,
the mirrors can be dispensed with altogether.
Choosing an appropriate ratio $|k_2|/|k_3|$, the light pulse simply
returns to its starting point because of the topology of the space manifold.
\bibitem{endnoteCPLEAR}
A different behavior of the clocks $C$ and $C^\prime$
would certainly be a more direct observation of
time reversal (``motion reversal'') noninvariance
than the recent result reported by the  CPLEAR Collaboration,
A. Angelopoulos et al., Phys. Lett. {\bf B 444}, 43 (1998).
%``First direct observation of time-reversal non-invariance in the  neutral kaon system,''
%%CITATION = PHLTA,B444,43;%%
But, for the moment, these clocks remain a \emph{Ge\-dan\-ken\-ex\-per\-i\-ment}.
The main practical difficulty is to keep the two pulses narrow enough to show
the effect, which is assumed to be nonvanishing ($m\sim \alpha/L \neq 0$).
For light pulses with
$|k_3|\sim 2\pi/L \ll |k_2|\equiv 2\pi/\overline{\lambda}$, a rough estimate
[using $v_{\mathrm g}/c \approx 1 - (1/8)\,m^2/\,k_2^2\,$]
gives the condition
$c\, T_{\mathrm exp} \gg \alpha^{-4}\,L^2/\overline{\lambda}$,
where $T_{\mathrm exp}$ is the total duration of the experiment
($T_{\mathrm exp} \approx N\, 2 D/c$, for a light pulse making $N$ round trips).
The experiment, which consists of clocks $C$ and $C^\prime$ of Fig.~1 pushed
against each other and synchronized initially on the mirrors touching,
would then have to run for a very long time ($\gg 10^{10}\, {\mathrm years}$),
because the \emph{present} universe is known to be very big
($L \gtrsim 10^{10}\, {\mathrm lightyears}$).
The size of the experiment,  $2D \approx c\,T_{\mathrm exp}/N$, should
also be sufficiently large, in order to reduce the total energy loss
of each pulse (the total number of reflections being $2N$).
On the other hand, the reflections in the clocks $C$ and $C^\prime$
are essentially the same (by a rotation of $180^\circ$),
which would, in principle,
make for a clean effect $N\,\Delta t^\prime \neq N\,\Delta t$, if it were there.
Also, this time difference $N\,(\Delta t^\prime -\Delta t)$ would be much larger
than the Planck time $t_{\mathrm P} \equiv (\hbar \,G/c^5)^{1/2}$,
provided the average wavelength is sufficiently large,
$ \overline{\lambda} \gg \alpha \,c \,t_{\mathrm P}$.
\bibitem{AN70}
A. Aharony and Y. Ne'eman, Int. J. Theor. Phys. {\bf 3}, 437 (1970).
\bibitem{endnoteKS}
More precisely, the spacetime manifold should have at least one separable compact
spatial dimension (coordinate $x^3$) with periodic spin structure;
see Ref. \cite{K00}. An example in the present context would be the
spatially homogeneous Kantowski--Sachs model with
spacetime topology ${\mathbb R} \times S^2  \times S^1 $, which has both
closure in time (i.e., recollapse after a period of expansion) and
CPT violation (at least, for an appropriate chiral gauge theory).
The spatially homogeneous and isotropic ``flat'' Friedmann--Robertson--Walker
(FRW) model with spacetime topology ${\mathbb R}^3\times S^1$,
on the other hand, expands forever (the same holds for the flat FRW model with the
topology ${\mathbb R} \times S^1 \times S^1 \times S^1$).
For further details on the topology of cosmological models, see, e.g.,
I. Ciufolini and J. A. Wheeler, \emph{Gravitation and
Inertia} (Princeton University Press, Princeton, NJ, 1995), Sec. 4.3.
\bibitem{endnoteFixedn}
The ticks from Eqs. (\ref{Dt}) and (\ref{Dtprime}) differ because of a
crucial sign difference in the ${\mathrm O}(m^3)$ term.
This remains so even if the size $L$ of the compact dimension
becomes time dependent [cf. Eq. (\ref{m})],
provided the odd integer $n$ (which traces back to the ultraviolet
regularization) is fixed once and for all.
\end{references}
\end{document}